\documentclass[11pt]{article}
\usepackage[T1]{fontenc}
\usepackage{lmodern}
\usepackage{amsmath,amssymb,amsthm}
\usepackage[margin=1in]{geometry}
\usepackage[colorlinks=true,linkcolor=blue,citecolor=blue,urlcolor=blue]{hyperref}

\newtheorem{theorem}{Theorem}
\newtheorem{lemma}[theorem]{Lemma}
\newtheorem{corollary}[theorem]{Corollary}
\newtheorem{proposition}[theorem]{Proposition}
\theoremstyle{remark}
\newtheorem{remark}[theorem]{Remark}

\newcommand{\bits}{\{0,1\}}
\newcommand{\F}{\mathbb{F}}
\newcommand{\C}{\mathbb{C}}
\newcommand{\Z}{\mathbb{Z}}
\newcommand{\poly}{\mathrm{poly}}
\newcommand{\BPTIME}{\mathrm{BPTIME}}
\newcommand{\BQTIME}{\mathrm{BQTIME}}
\newcommand{\PostBPTIME}{\mathrm{PostBPTIME}}
\newcommand{\PostBQTIME}{\mathrm{PostBQTIME}}
\newcommand{\NTIME}{\mathrm{NTIME}}
\newcommand{\DSPACE}{\mathrm{DSPACE}}
\newcommand{\DTIME}{\mathrm{DTIME}}
\newcommand{\PPTIME}{\mathrm{PPTIME}}

\title{Randomized, Quantum and Postselected Multitape Time\\
in Deterministic Space $t/\sqrt{\log t}$}
\author{Sebastian Ben Daniel}
\date{November 2026}

\begin{document}
\maketitle

\begin{abstract}
We show that a bounded-error multitape Turing machine running in time $t$
can be simulated deterministically in space $O(t/\sqrt{\log t})$, with no
constructibility assumption on $t$. The simulation computes any bit of the
exact number of accepting coin strings, so it also covers majority,
postselected, nondeterministic and modular acceptance. It uses the
tensor-network contraction behind the recent refutations of the Exponential
Time Hypothesis, which Cruz used to put nondeterministic time $T$ in space
$T(\log\log T)^2/\sqrt{\log T}$. Our space accounting removes the
$(\log\log T)^2$ factor.

Writing a pair of quantum paths as one weighted classical run, we obtain the
same bound for bounded-error quantum time with polynomial-time computable
amplitudes, for postselected quantum time with algebraic amplitudes, and for
quantum Merlin--Arthur time with witnesses of $O(t/(\log t)^{3/2})$
qubits. For explicit circuits no tape geometry is involved: acceptance
probabilities of $S$-gate quantum circuits, and $\#$CircuitSAT for $S$-gate
Boolean circuits, are computed exactly in space $O(S/\sqrt{\log S})$ plus the
logarithm of the description length.

As consequences, for every constant $K$ some language in deterministic
linear space differs, at every large input length, from every bounded-error
or postselected quantum machine running in time $o(n\sqrt{\log n})$ with $Kn$
bits of advice; another has correlation $O((\log n)/\sqrt n)$ with every
randomized machine and every quantum machine with algebraic amplitudes that
runs in that time with $\log^2n$ bits of advice, with or without a
bounded-error promise; and $\DSPACE(n)$ is not contained in
$\BPTIME[o(n\sqrt{\log n})]$ with constant-length advice that may depend on
the coins. Finally, exact counting or postselected simulation in space
$T^{1-\varepsilon}$ would refute the Weak Exponential Time Hypothesis, so a
polynomial improvement for bounded-error time has to use the bounded-error
promise.
\end{abstract}

\section{Introduction}

Williams~\cite{Wil25} proved that deterministic multitape time $t$ fits in
space $O(\sqrt{t\log t})$, using the tree evaluation algorithm of Cook and
Mertz~\cite{CM24}; this improved the $t/\log t$ bound of Hopcroft, Paul and
Valiant~\cite{HPV77}. Neither argument applies to bounded-error or quantum
machines. Both recompute pieces of the computation, and a randomized machine
would have to see the same coins at every recomputation. Enumerating coin
strings gives $\BPTIME[t]\subseteq\DSPACE(t)$, and enumerating pairs of
computation paths gives the same linear bound for quantum time. Under a
circuit lower bound for polynomial space, Santhanam put polynomially bounded
randomized time $t$ in space $t/\log t$~\cite[Theorem~21]{San02}. As far as
we know, no unconditional bound of the form $o(t)$ was known for
bounded-error time before the developments described next.

In October 2026 OpenAI~\cite{Ope26} and Anthropic~\cite{Cla26} posted
algorithms that count satisfying assignments of $k$-CNF formulas in
subexponential time; if correct, they refute the Exponential Time Hypothesis.
Both contract Boolean tensor networks, and Kintali gave a simplified
exposition of the OpenAI argument~\cite{Kin26}. The counting algorithm has
since been applied to machines and circuits:
\begin{itemize}
\item Strassle and Subercaseaux~\cite{SS26} proved
      $\NTIME[t]\subseteq\DTIME[2^{O(t/\sqrt{\log t})}]$.
\item Schlesinger~\cite{Sch26} combined the counting algorithm with
      Hopcroft--Paul--Valiant blocking and put $\NTIME[T]$, $\BPTIME[T]$ and
      $\PPTIME[T]$ in deterministic time $2^{O(T/\sqrt{\log T})}$ with space
      polynomial in $T$. The Lean library in which this was formalized also
      proves the same time bound, with polynomial space, for $\BQTIME[T]$,
      $\PostBQTIME[T]$ and related classes defined by uniform circuit
      families over Hadamard and Toffoli gates~\cite{CSL26}.
\item Tulsyan~\cite{Tul26} computed exact amplitudes of $G$-gate
      Clifford$+T$ circuits in time $2^{O(G/\sqrt{\log G})}$, assuming
      results of~\cite{Cla26}.
\item Cruz~\cite{Cru26} implemented Kintali's form of the counting algorithm
      in small space and proved
      $\NTIME(T)\cup\mathrm{coNTIME}(T)\subseteq\DSPACE(\kappa(T))$ with
      $\kappa(T)=T(\log\log T)^2/\sqrt{\log T}$, hence
      $\NTIME(n)\subsetneq\DSPACE(n)$. His draft states that the same
      separations hold for probabilistic and parity time.
\end{itemize}
Only Cruz's work gives a sublinear space bound, and the idea of counting
accepting computations in sublinear space with this contraction is his. This
note sharpens the bound, extends it to bounded-error, postselected and
quantum machines and to explicit circuits, and draws the consequences for
$\DSPACE(n)$. Table~\ref{tab:results} summarizes the bounds.

\begin{theorem}\label{thm:main}
Let $M$ be a multitape Turing machine with a two-way read-only input tape, an
optional read-only advice tape, and at most one fresh fair coin per step.
There is a deterministic algorithm that, given $x$, the advice, a binary time
cap $T\ge2$ and an index $0\le i\le T$, outputs bit $i$ of the number of
accepting strings of $T$ step-choice bits in work space
$O_M(T/\sqrt{\log(T+2)})$. Deterministic steps, and steps after a halt,
ignore their choice bit. In particular, for every diverging worst-case time
bound $t$ and every advice length $a$,
\[
 \BPTIME[t]/a\ \subseteq\ \PPTIME[t]/a\ \subseteq\
 \DSPACE\!\left(\frac{t}{\sqrt{\log(t+2)}}\right)/a,
\]
and the same inclusion holds for $\PostBPTIME[t]/a$, $\NTIME[t]/a$ and
$\mathrm{Mod}_m\mathrm{TIME}[t]/a$.
\end{theorem}

\begin{theorem}\label{thm:quantum-main}
Let $M$ be a multitape quantum Turing machine, with optional intermediate
measurements of its control state (Section~\ref{sec:model}), and let $x$, the
advice and a binary time cap $T\ge2$ be given.
\begin{enumerate}
\item If the amplitudes of $M$ are algebraic, lying in a fixed number field
      $\mathbb K$ with a fixed integral basis, then any bit of the
      coordinates of the scaled acceptance and postselection weights
      $Z_{\mathrm{acc}},Z_{\mathrm{post}}\in\mathcal O_{\mathbb K}\cap\mathbb
      R$ can be computed, and the comparisons $rZ_{\mathrm{acc}}\ge jD^{2T}$
      (for integers $0\le j\le r\le\poly(T)$) and $2Z_{\mathrm{acc}}\ge
      Z_{\mathrm{post}}$ decided, in work space $O_M(T/\sqrt{\log(T+2)})$.
\item If the real and imaginary parts of the amplitudes are
      polynomial-time computable, then under the bounded-error promise the
      acceptance of $M$ is decided in the same space, and without a supplied
      cap in space $O_M(t/\sqrt{\log(t+2)})$, where $t$ is the worst-case or
      the expected running time on $x$.
\end{enumerate}
In particular, for every diverging time bound $t$ and every advice length $a$,
\[
 \BQTIME[t]/a\,\cup\,\PostBQTIME[t]/a\ \subseteq\
 \DSPACE\!\left(\frac{t}{\sqrt{\log(t+2)}}\right)/a ,
\]
the first with polynomial-time computable amplitudes, the second with
algebraic amplitudes.
\end{theorem}

Part~(2) matters for fine-grained time: compiling computable amplitudes into
a fixed finite gate set costs a polylogarithmic factor per step, which would
cancel the $\sqrt{\log t}$ saving.

\begin{corollary}\label{cor:qma}
A quantum Merlin--Arthur verifier with algebraic amplitudes,
running in time $T$ with a $w$-qubit witness ($w\le T$), is simulated in
deterministic space $O(T/\sqrt{\log(T+2)}+(w+\log(T+2))\log(w+2))$. This is
$O(T/\sqrt{\log T})$ for $w=O(T/(\log T)^{3/2})$, and $o(T)$ for
$w=o(T/\log T)$.
\end{corollary}

For circuits the wiring is given, so no boundary data has to be enumerated.

\begin{theorem}\label{thm:circuits}
Fix a finite set of one- and two-qubit gates with algebraic entries,
together with single-qubit measurements in the computational basis.
\begin{enumerate}
\item Given an explicit circuit of $S\ge2$ such gates, of description length
      $L$, with prescribed initial basis values on its qubits and prescribed
      final values $\tau$ on a set of designated qubits, any bit of the
      coordinates of $D^{2S}\Pr[\text{the designated qubits read }\tau]$ in a
      fixed integral basis ($D$ clears the denominators of the gate entries,
      Section~\ref{sec:circuits}), and whether this probability is at least $j/r$
      for given integers $0\le j\le r\le\poly(S)$, or at least $j/r$
      times the probability of another such event, can be computed in
      deterministic space $O(S/\sqrt{\log(S+2)}+\log(L+2))$.
\item Given an explicit fan-in-$2$ Boolean circuit $C(x,y)$ with $S\ge2$
      gates and unrestricted fan-out, of description length $L$, and a
      string $x$, any bit of $|\{y:C(x,y)=1\}|$ can be computed in the same
      space.
\end{enumerate}
For circuits produced gate by gate by a generator that runs in space $s$,
the bounds hold with $s$ in place of $\log(L+2)$.
\end{theorem}

\begin{corollary}\label{cor:intro-hierarchy}
Unconditionally:
\begin{enumerate}
\item for every constant $K$ there is a language $L_K\in\DSPACE(n)$ that
      agrees on only finitely many input lengths with each language in
      $\PostBQTIME[t]/(Kn)$, $\BQTIME[t]/(Kn)$, $\PostBPTIME[t]/(Kn)$ and
      $\BPTIME[t]/(Kn)$ for $t=o(n\sqrt{\log n})$, and with each language
      decided with bounded error or with postselection by logspace-uniform
      quantum circuits of size $o(n\sqrt{\log n})$ with $Kn$ bits of advice;
\item there is a language $L^*\in\DSPACE(n)$ such that every quantum
      (algebraic amplitudes) or randomized multitape machine running in time
      $o(n\sqrt{\log n})$, and every logspace-uniform quantum circuit family
      of that size, with
      at most $\log^2n$ bits of advice and no bounded-error promise, agrees
      with $L^*$ on a uniformly random input of length $n$ with probability
      $\frac12\pm O((\log n)/\sqrt n)$, for all large $n$;
\item with unrestricted coins and pointwise tape-dependent advice, $\DSPACE(n)$
      is contained neither in $\BPTIME[o(n\sqrt{\log n})]//_{\mathrm{pt}}O(1)$
      nor, for any $\varepsilon>0$, in
      $\BPTIME[O(n)]//_{\mathrm{pt}}((\frac12-\varepsilon)\log\log n)$.
\end{enumerate}
\end{corollary}

Corollary~\ref{cor:intro-hierarchy} is proved in
Section~\ref{sec:separations} (Theorem~\ref{thm:separations} and
Corollary~\ref{cor:pt-advice}).

\begin{table}[t]
\centering
\small
\renewcommand{\arraystretch}{1.25}
\begin{tabular}{@{}lll@{}}
\hline
Model, time $t$ or $S$ gates & Previous space bound & This note\\
\hline
$\NTIME[t]$, $\mathrm{coNTIME}[t]$ & $t(\log\log t)^2/\sqrt{\log t}$~\cite{Cru26}
 & $t/\sqrt{\log t}$\\
$\BPTIME[t]$, $\PPTIME[t]$, $\PostBPTIME[t]$ & $t$; $\kappa(t)$ stated in~\cite{Cru26}
 & $t/\sqrt{\log t}$\\
$\BQTIME[t]$, computable amplitudes & $t$ & $t/\sqrt{\log t}$\\
$\PostBQTIME[t]$, algebraic amplitudes & $t$ & $t/\sqrt{\log t}$\\
QMA time $t$, $w$-qubit witness & --- & $t/\sqrt{\log t}+w\log w$\\
$S$-gate quantum circuits, $\#$CircuitSAT & $S$ & $S/\sqrt{\log S}$\\
\hline
\end{tabular}
\caption{Deterministic space bounds, up to constant factors and additive
logarithmic terms. ``$t$'' and ``$S$'' are the bounds obtained by
enumerating coin strings, paths or assignments. Cruz's draft states the
separations for probabilistic and parity time without the details. For
circuits and for quantum time classes the time bound $2^{O(S/\sqrt{\log S})}$
with polynomial space is in~\cite{Tul26,CSL26}.}
\label{tab:results}
\end{table}

\paragraph{What is new.} Compared with~\cite{Cru26}:
\begin{enumerate}
\item The space bound loses no $(\log\log T)^2$ factor, so the separation from
      linear space is at time $o(n\sqrt{\log n})$ rather than
      $o(n\sqrt{\log n}/(\log\log n)^2)$.
\item Bounded-error, majority and postselected acceptance are stated and
      proved explicitly, the algorithm outputs every bit of the exact count,
      and advice is carried along.
\item The simulation covers bounded-error quantum time with computable
      amplitudes, postselected quantum time and quantum Merlin--Arthur time
      with algebraic amplitudes, and explicit quantum and Boolean circuits
      (Theorem~\ref{thm:circuits}). Time bounds of the form
      $2^{O(S/\sqrt{\log S})}$ for exact Clifford$+T$
      amplitudes~\cite{Tul26} and for quantum time classes over a fixed gate
      set~\cite{CSL26} were known with polynomial space; they also follow
      from Theorem~\ref{thm:circuits}. For quantum machines and circuits,
      what is new is the space bound.
\item The argument needs no block-respecting normal form and no
      constructible clock. Epochs declare their whole accessible tape
      interval (Lemma~\ref{lem:epochs}), and the time cap is found by
      doubling (Section~\ref{sec:reconstruction}).
\item The separations of Corollary~\ref{cor:intro-hierarchy}(2) and~(3),
      on average over the uniform distribution and against pointwise
      tape-dependent advice, are new even for randomized time, as far as we
      know.
\end{enumerate}
The main technical point is the space accounting. The preprint~\cite{Cla26}
counts operations and does not claim a space bound; its construction of the
crossing identities (Appendix~B there) builds tables of size exponential in
the block length. We apply those identities only to blocks of
$b=\lfloor(\log T)^{3/4}\rfloor$ bits. Each table then has $2^{O(b)}=T^{o(1)}$
bits, while the labels that select one identity per block cost a fraction
$O(\log b/b)=o(1/\sqrt{\log T})$ of the network width. The rest of the
accounting (Lemma~\ref{lem:clique}) checks that the nested table queries of
the recursion never copy their arguments: each argument is a view of
counters already held by some caller. The quantum case needs no new network:
a pair of quantum paths is a run of a classical nondeterministic machine
whose transitions carry the amplitudes of the two paths as weights
(Lemma~\ref{lem:quantum-paths}), and the contraction runs unchanged over a
finite quotient of the ring of integers of the amplitude field.

\paragraph{Limits of exact counting.} Theorems~\ref{thm:main}
and~\ref{thm:quantum-main} compute exact counts, so they also decide
nondeterministic time. Proposition~\ref{prop:limits} shows that this caps the
method: exact counting, and even bounded-error postselected simulation, in
space $T^{1-\varepsilon}$ for a constant $\varepsilon>0$ would refute the
Weak Exponential Time Hypothesis, which the MIT Hardness Group recently
showed to be equivalent to the Stearns--Hunt Satisfiability
Hypothesis~\cite{MIT26}. A simulation of bounded-error time in space
$t^{1-\varepsilon}$, let alone near Williams's $\sqrt t$, has to use the
bounded-error promise and the absence of postselection. The companion
paper~\cite{BDone} gives further barriers for the tree-evaluation version of
this question and unconditional $O(\sqrt T\log T)$ bounds for machines with
one active work tape.

\paragraph{Status of the ingredients.} The algebraic input, the grouping
lemma and the crossing identities with $\rho=1/\sqrt2$, is taken from the
Anthropic preprint~\cite[Lemma~8, Definition~7, Theorems~A and~B]{Cla26}.
The preprint is not refereed, and as far as we know no person has checked
its proofs. Its Lean~4 development has counterparts of Lemma~8, of Theorem~B
and of the two operator lemmas behind Theorem~B. The time clause of
Definition~7 is formalized only in a weak form, so the bound of $2^{O(m)}$
work bits for the pieces of length $m$, which we use, rests on the printed
argument in~\cite[Appendix~B]{Cla26}; we have checked that argument. We have
also checked the identity $\sum_iA_{m,i}=d_m[x=v][y=u]$ of Theorem~B by
exhaustive computation for all partitions of $h\le7$, and the main-text
pieces with $\rho=11/12$ (Remark~\ref{rem:rho}) by hand. Everything else
below is ours and is not formalized. If the crossing
identities were available only with a weaker constant $\rho<1$, all results
would hold with space $T/(\log T)^{\log_2(1/\rho)}$, still $o(T)$;
Remark~\ref{rem:rho} makes this precise.

\paragraph{Organization.} Section~\ref{sec:model} fixes the machine models.
Section~\ref{sec:networks} turns the accepting coin strings of a machine into
a sum of network values. Sections~\ref{sec:grouping} to~\ref{sec:recursive}
evaluate one network in space $O(T/\sqrt{\log T})$; this is where the space
accounting happens. Section~\ref{sec:reconstruction} recovers the exact count
and proves Theorem~\ref{thm:main}. Section~\ref{sec:quantum} treats quantum
machines, Section~\ref{sec:circuits} explicit circuits,
Section~\ref{sec:separations} the separations from linear space, and
Section~\ref{sec:limits} the limits of exact counting.

\section{Models}
\label{sec:model}

We use a fixed finite number of work tapes, a read-only two-way input tape
with endmarkers, optionally a read-only two-way advice tape, and head motions
of at most one cell per transition. A deterministic simulator has a fixed
finite alphabet. Logarithms have base two unless indicated otherwise.

\paragraph{Classical machines.} Coins are fresh independent fair bits. Fix a
machine $M$, an input $x$ with its advice, and a time cap $T\ge2$. Give every
clock step a choice bit, ignored whenever the transition is deterministic.
After halting, the outcome, heads and tapes stay stationary through step $T$.
For a set $\mathcal C$ of control states write $A_{\mathcal C}(x,T)$ for the
number of strings in $\bits^T$ whose run is in $\mathcal C$ at step $T$, and
$A(x,T)$ for the accepting ones; the probability of accepting by the cap is
$2^{-T}A(x,T)$. A postselected machine~\cite{Aar05} has a set of
postselecting halting states containing the accepting ones, reaches it with
positive probability, and accepts with conditional probability at least
$\frac23$ or at most $\frac13$; this defines $\PostBPTIME[t]$.

\paragraph{Quantum machines.} A multitape quantum Turing machine in the
sense of Bernstein and Vazirani~\cite{BV97} has, for each local situation
(control state and scanned symbols), at most $\Delta=O_M(1)$ possible
transitions $e$, each with an amplitude $\alpha_e$, and the resulting step
operator $U_x$ on basis configurations is unitary. After every step the
machine measures a function $m$ of its control state in the computational
basis (a scanned cell can be measured by first copying it into the control).
We require that $m(q)$ determine whether $q$ is a halting state, and a branch
in which a halting state has been measured is left unchanged from then on.
We treat this as a stationary transition with amplitude $1$. Without the
measurement an absorbing halting state would not be unitary, since a halting
configuration would be reached both from itself and from the configuration
that halted into it. With it, one step is the trace-preserving operation
$\rho\mapsto\sum_\mu V\Pi_\mu\rho\,\Pi_\mu V^\dagger$, where $\Pi_\mu$
projects onto the configurations with measured value $\mu$, and $V$ acts as
$U_x$ on nonhalting and as the identity on halting configurations; each
$V\Pi_\mu$ is an isometry on the range of $\Pi_\mu$. Writing the
measurement before $V$ rather than after it changes nothing we use: the
initial state is a basis state, so the first measurement acts trivially, and
the final probabilities are those of diagonal projectors, which commute with
every $\Pi_\mu$. Synchronously halting
machines~\cite{BV97} are the case in which $m$ records only halting.

The amplitudes are of one of two kinds.
\emph{Algebraic:} all $\alpha_e$ lie in a fixed number field
$\mathbb K\subset\C$ closed under complex conjugation, and $D\ge1$ is an
integer with $D\alpha_e\in\mathcal O_{\mathbb K}$ for all $e$, including the
stationary transitions. Fix an integral basis $\gamma_1,\dots,\gamma_d$ of
$\mathcal O_{\mathbb K}$. With $\mathbb K=\mathbb Q(\zeta_8)$,
$\zeta_8=e^{\mathrm i\pi/4}$, and $D=2$ this covers Clifford$+T$ and
Hadamard--Toffoli; $D=10$ covers the rational amplitudes used
in~\cite{BV97}. \emph{Computable:} for every $e$ and $k$, dyadic
approximations of $\operatorname{Re}\alpha_e$ and $\operatorname{Im}\alpha_e$
within $2^{-k}$ are computable in time $\poly(k)$. This is the amplitude set
of~\cite{BV97}.

For algebraic amplitudes a path $p$ of $T$ steps from the initial
configuration $c_0$ has weight $W(p)=\prod_tD\alpha_{e_t}\in\mathcal O_{\mathbb
K}$. The probability that the configuration at step $T$ has control state in
a set $\mathcal C$ is then $D^{-2T}Z_{\mathcal C}(x,T)$, where
\[
 Z_{\mathcal C}(x,T)=\sum_{c}\ \sum_{\substack{p_1,p_2:c_0\to c\\
 m(p_1(t))=m(p_2(t))\ \forall t\le T}}W(p_1)\overline{W(p_2)},
\]
the outer sum ranging over configurations $c$ with control state in
$\mathcal C$. Exchanging $p_1$ and $p_2$ conjugates each term, so
$Z_{\mathcal C}\in\mathcal O_{\mathbb K}\cap\mathbb R$. The machine runs in
time $t$ if the probability of a nonhalting control state at step $t(|x|)$
is $0$. $\BQTIME[t]$ is defined by the thresholds $\frac23$ and $\frac13$
with computable amplitudes, and $\PostBQTIME[t]$ by the conditional
thresholds of~\cite{Aar05} with algebraic amplitudes; postselection on
exponentially small events does not tolerate approximate amplitudes. For
quantum Merlin--Arthur time, the
verifier starts with a $w$-qubit witness on cells $0,\dots,w-1$ of work tape
$1$; for $u,v\in\bits^w$ let $Z(u,v)$ be the same sum over accepting $c$ with
$p_1$ starting from witness $v$ and $p_2$ from witness $u$, so that
$Z(u,v)=D^{2T}\langle u|E_x|v\rangle$ for the acceptance operator $E_x$.

\section{From coin strings to networks}
\label{sec:networks}

\begin{lemma}\label{lem:epochs}
For $1\le g\le T$ and every set $\mathcal C$ of control states,
$A_{\mathcal C}(x,T)$ is the sum of the values of a uniformly
enumerable family of networks on exactly $g$ vertices. Each has
$P=O_M(\lceil T/g\rceil)$ and $U\le gP=O_M(T)$ (these parameters are defined
in Section~\ref{sec:grouping}). Its vertex weights $w_a$ are Boolean, and all
edge factors test equality of strings. The current network is specified by
$O_M(g\log(T+2))$ bits. With this specification available for reading, its
port lists and edge widths can be generated with $O_M(\log(T+2))$ auxiliary
bits, and its local predicates can be evaluated with
$O_M(\lceil T/g\rceil+\log(T+2))$ auxiliary bits.
\end{lemma}

\begin{proof}
Use epoch boundaries $t_a=\lfloor aT/g\rfloor$, $0\le a\le g$. Each epoch has
at most $b_0=\lceil T/g\rceil$ transitions. Enumerate the finite-control
state and the positions of every head, including the input and advice heads, at each
boundary. Fix the prescribed initial data and require the final state to lie
in $\mathcal C$. Head positions use $O(\log(T+2))$ bits. Use a unique encoding of
each such boundary list. Consistency between epochs through earlier visits
to the same cells follows the segmented multitape reduction
of~\cite{SS26}; here an epoch declares its entire accessible interval, which
removes the need for a block-respecting transformation.

At vertex $a$, for each work tape $j$, take the interval $I_{aj}$ of cells at
distance at most $b_0$ from its enumerated starting head position,
intersected with the legal tape domain. The setting $z_a\in S_a$ contains the
work symbols in all these intervals before and after the epoch, together with
exactly $t_a-t_{a-1}$ choice bits. Use fixed-width symbol encodings and reject
invalid codes. Given $z_a$, the predicate $w_a$ starts with its declared
initial symbols and the enumerated boundary data, simulates the indicated
transitions, and checks the resulting state, head positions, and entire final
arrays. It reads the true input $x$ and the advice through the simulated read-only
heads. No work
head can leave its interval during these steps. For a cell whose interval
occurrence is its first on that work tape, $w_a$ also requires its declared
initial symbol to be blank.

For each physical cell on each work tape, list the epochs whose interval
contains it in chronological order. Between successive epochs in that list
impose equality of the earlier final symbol and the later initial symbol.
Concatenate all tests between a pair of vertices in a fixed order, for
example by tape, cell, then symbol-bit index. If the concatenated string has
$c_{a,a'}$ bits, its equality test factors through exactly this many bits by
summing over a common string and testing equality at both ends. Each
occurrence has at most one incoming and one outgoing such comparison. Thus
both the setting length and the total port width at any vertex are at most
$C_Mb_0$ for a constant independent of $g$. With $P=C_Mb_0$ we have
$U=2\sum_{a<a'}c_{a,a'}\le gP=O_M(T)$, since $gb_0\le T+g$. The port list, its
widths, and the first-occurrence tests are generated by rescanning the
boundary list with tape, cell, and epoch counters. The local simulation needs
only its interval arrays and counters.

Every accepting choice string determines one boundary list and one tuple of
settings of weight one. Conversely, concatenate the local choice strings of a
passing tuple. Induction over epochs gives its actual computation. Every
starting cell is fixed either by its first blank check or by the final symbol
of the preceding interval containing that cell. Requiring the entire final
arrays also fixes the cells untouched during an epoch. This establishes the
claimed bijection, including runs that halt early or make fewer than $T$ coin
requests.
\end{proof}

\paragraph{Weighted runs.} Suppose every transition $e$ of $M$ carries a
weight $\beta_e$ in a commutative ring $R$, taken from a table $\beta$ with
$O_M(1)$ entries, and give a choice string the product of the weights of the
transitions in its run. Then the same construction expresses the weighted
count $A^\beta_{\mathcal C}(x,T)$, the sum of these products over the choice
strings whose run is in $\mathcal C$ at step $T$, as a sum of network values.
The only change is that a passing setting of vertex $a$ has weight the
product of the $\beta_e$ of the transitions it simulates, rather than $1$.
The bijection above preserves weights, because the weight of a run is the
product of the weights of its epochs. If the table entries are stored in
$s_\beta$ bits, a vertex weight is evaluated in
$O_M(\lceil T/g\rceil+\log(T+2))+O(s_\beta)$ auxiliary bits.

\section{Grouping local constraints}
\label{sec:grouping}

Sections~\ref{sec:grouping} to~\ref{sec:recursive} evaluate one network.
Grouping its $g$ vertices into $K$ groups turns it into a clique on $K$
vertices (Lemma~\ref{lem:group}). The grid reduction on short blocks turns
that clique, term by term, into a network on the rows (or columns) of a
grid, which grouping turns into a smaller clique, and so on
(Lemma~\ref{lem:clique}). A table entry at one level is a sum over the
settings of a group, and each term calls the table entries of the level
below; these calls go back down to the original tables and never invoke the
clique solver again, so the solver frames and the entry frames form one
chain whose length is the number of levels.

Fix a prime $p$ of $O(\log(T+2))$ bits. All arithmetic for a single network
evaluation takes place in $\mathbb F_p$. Sections~\ref{sec:grouping}
to~\ref{sec:recursive} use only addition, multiplication, and division by
integers prime to $p$. They therefore hold verbatim over any commutative ring
$R_p$ in which those integers are invertible and whose elements take
$O_R(\log p)$ bits. We use $R_p=\mathcal O_{\mathbb K}/p\mathcal O_{\mathbb K}$,
which is $(\Z/p)^d$ as an additive group in the integral basis; it need not be
a field.

A $(K,\lambda)$-clique has one $\lambda$-bit string on each unordered edge of
the complete graph $K_K$. A vertex table $W_v$ takes the $K-1$ incident
strings in a prescribed order. Its value is
\[
 Z=\sum_{(x_e)_e}\prod_v W_v\bigl((x_e)_{e\ni v}\bigr).
\]
The tables are procedures. The argument of a query is available for
read-only bit access inside the procedure, and may be specified by offsets
into counters that are already on the simulation tape. Suppose the original
table procedure needs $W$ further work bits. Throughout, $K$, $\lambda$, $p$
and all string lengths under consideration are bounded by fixed powers of a
common parameter $T\ge2$.

A network on $g$ vertices has settings $z_a$ of lengths $P_a\le P$, local
weights $w_a(z_a)$ and pair factors
\[
 \Psi_{ab}(z_a,z_b)=\sum_{\theta\in\{0,1\}^{c_{ab}}}
 f_{ab}(z_a,\theta)\,f_{ba}(z_b,\theta).
\]
Forbidden settings get weight zero. Here $c_{ab}=c_{ba}$,
$\sum_{b\ne a}c_{ab}\le P$, and $U=\sum_a\sum_{b\ne a}c_{ab}$. The value sums
the product of all vertex weights and all unordered pair factors over the
settings. The given procedures evaluate $w,\Psi,f$ modulo $p$. The cost of a
procedure is its additional space beyond its read-only arguments and the
network description.

We use the following grouping, which is what the proof of Lemma~8
of~\cite{Cla26} constructs.

\begin{lemma}[{\cite[proof of Lemma~8]{Cla26}}]\label{lem:grouping-import}
Let $K\ge2$ and $0<\zeta\le1$, and suppose that either $g=3K^8$ and
$\zeta=1/(2K)$, or $K$ divides $g$, $c_{ab}\le P/g$ for all $a\ne b$, and
$2\zeta^2g\ge K\ln(2gK)$. Then there is a map $\pi:[g]\to[K]$ such that every
group $\pi^{-1}(k)$ has at most $2g/K$ vertices and, for all $k\ne k'$,
\[
 \sum_{a\in\pi^{-1}(k)}\ \sum_{b\in\pi^{-1}(k')}c_{ab}
 \;\le\;\frac{U+2\zeta gP}{K^2}.
\]
\end{lemma}

\begin{lemma}\label{lem:group}
Under the hypotheses of Lemma~\ref{lem:grouping-import}, the network has the
value of a $(K,L)$-clique with
\[
 K^2L\le U+2\zeta gP+K^2 .
\]
If each original entry takes $W$ auxiliary bits, the new table entries take
at most $2gP/K+W$ auxiliary bits besides $O(g^2\log(T+2))$ bits of indices
and description.
\end{lemma}

\begin{proof}
Put $L=\max\{1,\lceil(U+2\zeta gP)/K^2\rceil\}$. Find a map $\pi$ as in
Lemma~\ref{lem:grouping-import} by enumerating maps $[g]\to[K]$ and checking
the two conditions. Order the old pairs between each group pair and use the
first bits of the new edge index for their respective $\theta$ strings.
Require zero on the padding. A group's entry sums over all settings of its
vertices. For each setting it multiplies the vertex weights, the $\Psi$
values of wholly internal pairs, and its endpoint $f$ on every crossing pair.
Expanding the grouped clique gives the stated network value.

One counter of at most $2gP/K$ bits enumerates the settings of the group. All
old weights and factors are called sequentially, using slices of this counter
and of the supplied edge indices. Group maps, pair widths and offsets fit in
the displayed description bound. The only arithmetic state is a constant
number of modular accumulators and loop indices.
\end{proof}

\section{The grid reduction on short blocks}

Write $\rho=1/\sqrt2$. For each length $m$, Definition~7 and Theorem~B
of~\cite{Cla26} give a positive integer $d_m$, a set $I_m$ of at most
$(2m)^c$ labels, and integer functions $A_{m,i}$ with
\[
 \sum_{i\in I_m}A_{m,i}(x,y,u,v)=d_m\,[x=v][y=u]
\]
for $m$-bit arguments $x,y,u,v$. For each label the function factors across
$(x,y)\mid(u,v)$ through $r_i$ bits and across $(x,u)\mid(y,v)$ through $s_i$
bits, in the sense of the pair-factor formula above, and
\[
 r_i,s_i\le2m,\qquad r_i+s_i\le4\rho m+c\log(2m).
\]
The individual caps hold because any matrix on two $2m$-bit arguments
factors through $2m$ bits, using one argument as the index.

\paragraph{Space for the building blocks.} The construction in the proof of
Theorem~B uses partitions of size $h=O(m^2)$, at most $2^{O(m)}$ table
entries, and integers of $O(h^2)$ bits in the table
calculation~\cite[Appendix~B]{Cla26}. It therefore generates labels, widths,
normalizing constants and factor entries in $2^{O(m)}$ work bits on an
ordinary Turing machine, and $\log(2d_m)=(2m)^{O(1)}$. Fix
\[
 \ell=\lceil\log(T+2)\rceil,\qquad b=\lfloor\ell^{3/4}\rfloor,\qquad
 \eta=\log(2b)/b .
\]
The integer $b$ is found by testing $b^4\le\ell^3$. Individual factor calls
on blocks of length at most $b$ take $2^{O(b)}=T^{o(1)}$ bits. This is the
only place where tables of the contraction are stored explicitly.

\paragraph{Admissible primes.} The proof of Theorem~B takes $d_m=h!$ with
$h=O(m^2)$, so every prime up to $h$ divides $d_m$. We call a prime
\emph{admissible} if it divides none of $d_1,\dots,d_b$; in particular no prime
up to the partition size $h$ used for $m=b$ is admissible. Since
$h=O(b^2)=O((\log T)^{3/2})$, testing admissibility amounts to skipping the
primes up to that size, and $d_m\bmod p$ is computed within the $2^{O(b)}$
work bits charged above. All
residues below are taken modulo admissible primes, and
Section~\ref{sec:reconstruction} checks that enough of them have
$O(\log(T+2))$ bits.

\paragraph{The grid.} Place the vertices of a $(g^2,\lambda)$-clique on a $g$
by $g$ grid and put $N=g^4\lambda$. Give each endpoint of each edge its own
$\lambda$-bit string and impose equality between the two copies. Every pair
of diagonals of a grid rectangle now imposes $[x=v][y=u]$ on its four
corresponding strings, listed in row order. Every edge joining corners in
different rows and different columns belongs to exactly one such diagonal
pair.

Split all four strings at the same boundaries into blocks of $b$ bits and a
final shorter block when needed. For each rectangle and block of length $m$,
select a label in $I_m$. The sum over all these choices, divided by the
product of the $d_m$'s (one per block per rectangle), restores the original
value modulo $p$. The denominator depends only on the block lengths and is
invertible modulo an admissible prime. Enumerate the choices one at a time.
Storing the ordinal of each selected label costs
\[
 O\bigl(N\eta+g^4\log(2b)\bigr)
\]
bits, since there are $\binom g2^2$ rectangles. For a selected term, the row
width $r$ and column width $s$ of a rectangle are the sums of the
corresponding block widths. Thus $r,s\le2\lambda$ and
\[
 r+s\le4\rho\lambda+O\bigl((\lambda/b+1)\log(2b)\bigr).
\]

A term is a network on the $g$ rows. One row setting consists of all endpoint
strings of its grid points and has length at most $P=g^3\lambda$. Its weight
multiplies the original tables and the straight equalities internal to the
row. Between a pair of rows there are $g$ straight equalities and
$\binom g2$ rectangle factors. Their indices concatenate, so the pair width
is at most $g\lambda+2\lambda\binom g2=g^2\lambda=P/g$. Exchanging rows and
columns gives a second network for the same term. In either orientation
$gP=N$. In the average of the two totals $U$, each rectangle contributes
$r+s$ (rows give the $(xy)\mid(uv)$ split and columns the $(xu)\mid(yv)$
split), and the straight edges contribute at most $N/g$. Consequently one
can choose an orientation with
\[
 U\le(\rho+1/g+C_1\eta)N+O(g^4\log(2b))
\]
for an absolute constant $C_1$. Both totals are computable from the ordinals
alone. To evaluate a pair factor without a split, multiply the small
$A_{m,i}$ values and test its straight equalities. To evaluate an endpoint,
multiply the small endpoint factors and test the straight-edge index slices.
All products are sequential. Any requested index slice is located by
rescanning the ordinal string and adding its widths, which avoids storing a
separate offset per short block.

\section{Recursive workspace}
\label{sec:recursive}

Use the sizes from Theorem~A(a) of~\cite{Cla26} with a sufficiently large
constant integer $x_0\ge18$:
\[
 K_0=2^{x_0},\qquad \zeta_j=2^{-j-5},\qquad
 g_j=2^{3j+x_0}K_j,\qquad K_{j+1}=g_j^2.
\]
Then $K_j\mid g_j$, $1/g_j\le\zeta_j$ and $2\zeta_j^2g_j\ge K_j\ln(2g_jK_j)$,
as one checks by substituting $\log K_j=(3x_0+6)2^j-6j-2x_0-6$. In particular
$\log K_j=\Theta(2^j)$ and $K_{j+1}\le K_j^{10}$.

\begin{lemma}\label{lem:clique}
Let $N=K_j^2\lambda$. Suppose $N$ is polynomially bounded in $T$ and the
original table procedures require $W$ bits besides their arguments. There
are constants $B,C$ such that the $(K_j,\lambda)$-clique can be evaluated
modulo every admissible prime in at most
\[
 (e_j+B\eta)N+W+O\bigl(K_j^C\log^2(T+2)+(j+1)2^{O(b)}\bigr)
\]
bits of stored data and scratch. Here $e_0=1/2$ and
$e_{i+1}=(\rho+3\zeta_i)e_i+3/K_i$, so $e_j=O(\rho^j)$. A fixed-alphabet
multitape implementation costs a constant times this bound.
\end{lemma}

\begin{proof}
At the base, enumerate the $\binom{K_0}2\lambda$ edge-assignment bits and use
the appropriate incident-edge subsequence for each vertex query. At step
$i+1$, write $g=g_i$, $K=K_i$ and use the grid reduction, followed by the
grouping of Lemma~\ref{lem:group}. The smaller clique has parameter
\[
 N'\le(\rho+3\zeta_i+C_1\eta)N+O(g^4\log(2b)+K^2).
\]
A grouped entry enumerates at most $2N/K$ setting bits. It needs its previous
table workspace once, and the grid solver retains at most
$C_2\eta N+O(g^4\log(2b))$ bits for its selected labels. After evaluating one
term, add it to the modular sum. At the end divide by the normalization using
modular inversion; its residue is the product of the local normalizers
modulo $p$.

For sufficiently large $T$, the coefficient of $N$ at this step is at most
\[
 (e_i+B\eta)(\rho+3\zeta_i+C_1\eta)+2/K_i+C_2\eta\ \le\ e_{i+1}+B\eta .
\]
Indeed $\rho+3\zeta_i<0.81$ and $e_i\le1$ by induction. Choose $B$ larger than
a fixed multiple of $C_1+C_2+1$, and then take $\eta$ sufficiently small. So
the new charge for the stored labels is absorbed at every level. Also
$\sum_i\zeta_i<\infty$ and $\sum_i\rho^{-i-1}/K_i<\infty$; dividing the
recurrence by $\rho^{i+1}$ and iterating gives $e_i=O(\rho^i)$.

The solver holds one selection at each active grid level. At its base the
counter over edge settings supplies entries of the innermost grouped clique.
A grouped entry retains one counter over its local settings while it calls
weights at the previous grid. These weights call the previous table
procedure on their incident strings, and the calls proceed back through the
levels toward the original tables. A table-entry routine performs its local
sums and products without invoking the clique solver. Thus the number of live
solver and entry frames is $O(j+1)$, and the allowance $W$ has coefficient one
throughout.

A bit of a query argument is obtained from a counter by translating its
offset through these frames. For a group, concatenate pairs in fixed order.
For a row or column, concatenate vertices, their incident edges, and the
short-block indices in fixed order. A view stores the operation, the group or
grid indices and a reference to the preceding view. Computing a width uses
the saved choices and the short-block algorithms, independently of the
argument values. Retrieval therefore ends at a counter that is already
stored. Repeated sequential scans implement these views on tapes. Each stage
uses a polynomial in its grid size times $\log^2(T+2)$ bits for maps, port
lists, loop indices and view descriptors, in addition to the counters and
labels already charged. The additive width error in $N'$ fits in the same
bound. For large $T$ the coefficient of $N$ in the bound for $N'$ is less than
$0.9$, and all additive width errors are polynomial in the largest grid size
times $\log(T+2)$, so all reachable widths are polynomial in $T$. Since grid
sizes grow at least by squaring, summing the polynomial overheads over all
stages costs $O(K_j^C\log^2(T+2))$ for a fixed $C$. At most $O(j+1)$
short-block workspaces of $2^{O(b)}$ bits cover generation and indexing.
Small values of $T$ are handled by exhaustive enumeration.
\end{proof}

\begin{remark}[A weaker crossing constant]\label{rem:rho}
The only property of $\rho$ used above, besides $\rho<1$, is
$\rho+3\zeta_0<0.81$. If the identities were available only with some
constant $\rho\in(1/\sqrt2,1)$, for instance through the elementary
baseline of~\cite[Lemma~2(c)]{Cla26} with $\rho=11/12$, replace $\zeta_j$ by
$2^{-j-x_1}$ for a constant $x_1$ with $\rho+3\cdot2^{-x_1}<1$, adjust $x_0$
accordingly, and replace $0.81$ by a constant below $1$. Then
$e_j=O(\rho^j)=O((\log T)^{-\log_2(1/\rho)})$, and Theorems~\ref{thm:main}
and~\ref{thm:quantum-main}, Corollary~\ref{cor:qma} and
Theorem~\ref{thm:circuits} hold with
$T/(\log T)^{\log_2(1/\rho)}$ in place of $T/\sqrt{\log T}$; for $\rho=11/12$ the exponent is
about $0.125$. We have not written out the modified constants in full.
\end{remark}

\section{Exact reconstruction for classical machines}
\label{sec:reconstruction}

Choose a fixed integer $D_1$ sufficiently large compared with $C$ and $16$.
For large $T$ let $j$ be the largest index with $K_j\le T^{1/D_1}$. Then
$K_j>T^{1/(10D_1)}$ and $\log K_j=\Theta(\log T)$. Use $g=3K_j^8\le T$ epochs
in Lemma~\ref{lem:epochs}. For each boundary list, the grouping of
Lemma~\ref{lem:group} (first case, $\zeta=1/(2K_j)$) yields a clique with
$N\le U+gP/K_j+K_j^2=O_M(T)$. Its extra grouped-setting counter has
$O_M(T/K_j)$ bits, and the initial local predicates take
$O_M(T/g+\log(T+2))$ bits. The boundary list and initial grouping
description take $O_M(g^2\log(T+2))$ bits. All three costs are
$o(T/\sqrt{\log T})$.

Apply Lemma~\ref{lem:clique} to that clique. From $\rho=2^{-1/2}$ and
$\log K_j=\Theta(2^j)$ we get $e_j=O(1/\sqrt{\log T})$. Also
$\eta=O(\log\log T/(\log T)^{3/4})=o(1/\sqrt{\log T})$. The additive costs
are $O(T^{1/2}\log^2(T+2))+T^{o(1)}$, after increasing $D_1$ if necessary.
Enumerating the boundary lists and accumulating their residues reuses this
space. So, for every set $\mathcal C$ of control states,
$A_{\mathcal C}(x,T)\bmod p$ is computed in space $O_M(T/\sqrt{\log(T+2)})$
for every admissible prime $p$ of $O(\log(T+2))$ bits. For a weighted machine
with weights in $R$, the residue of $A^\beta_{\mathcal C}(x,T)$ in $R_p$ is
computed in space $O_M(T/\sqrt{\log(T+2)})+O(s_\beta)$ in the same way, with
the constant now depending also on $R$.

\paragraph{Chinese remaindering.} The inadmissible odd primes divide
$\prod_{m\le b}d_m$, so there are at most
$\log_2\bigl(\prod_{m\le b}d_m\bigr)=b^{O(1)}=(\log T)^{O(1)}$ of them.
Hence, for every $R\le\poly(T)$, the first $R$ admissible odd primes all have
$O(\log(T+2))$ bits, by the usual bound on the $k$th prime; with $R=T+1$
their product exceeds $2^T$. Generate them by trial division, checking the
$d_m$'s one at a time. Then convert from this Chinese remainder
representation to binary. If $Q=\prod_ip_i$, $Q_i=Q/p_i$ and
$v_i=(Q_i\bmod p_i)^{-1}\bmod p_i$, the required integer is the least residue
modulo $Q$ of $\sum_i(A(x,T)\bmod p_i)\,v_iQ_i$. Each $v_i$ is computed modulo
the short prime $p_i$, and the rest needs only iterated multiplication,
iterated addition and integer division of polynomially many numbers of
polynomial length, which are in uniform $\mathrm{TC}^0$~\cite{HAB02} and hence
in deterministic logspace. We run this logspace conversion on the moduli and
residues as a virtual input. Each requested bit of the virtual input
is generated afresh. Its explicit length would be $O(T\log(T+2))$, so the
converter adds $O(\log(T+2))$ bits beyond the space for one residue query.
It yields any bit of $A_{\mathcal C}(x,T)$, including bit $T$ when every
choice string reaches $\mathcal C$. This proves the counting assertion of
Theorem~\ref{thm:main}.

\paragraph{No clock.} To dispense with a supplied worst-case clock, start at
$T=2$ and double. Count the choice strings whose run is still going at the
cap. This count is zero exactly when every computation has halted; test it
by checking whether its residues at the first $T+1$ admissible odd primes are
all zero. Once it is zero, compare the acceptance count with $2^{T-1}$ for
bounded error or strict majority, compare $2A_{\mathrm{acc}}$ with
$A_{\mathrm{post}}$ for postselection, compare with zero for
nondeterminism, and reduce modulo $m$ (by Horner's rule on the bits) for
$\mathrm{Mod}_m\mathrm{TIME}$. For a fixed machine with more than two
nondeterministic alternatives, a binary branching simulation has constant time
overhead. The peak cap is at most $2\max\{2,t_M(x)\}$ for the worst-case time
$t_M(x)$ of the binary-branching machine. For a cap shorter than the input,
the real input head supplies a requested symbol by scanning from the left
end, respecting endmarkers; only positions within distance $T$ of the start
can be requested, so this takes $O(\log(T+2))$ work bits, and the advice tape
is handled the same way. No clock constructibility or coin budget beyond the
running time is required. This proves Theorem~\ref{thm:main}.

\section{Quantum machines}
\label{sec:quantum}

A pair of quantum paths is the run of a classical nondeterministic machine
whose transitions carry the amplitudes of the two paths as weights.

\begin{lemma}\label{lem:quantum-paths}
Let $M$ be a quantum machine as in Section~\ref{sec:model} and $\mathcal C$ a
set of its control states. There are a constant $c_\Delta$ and a classical
multitape machine $M'$ with at most one binary choice per step, both
depending only on $M$ and $\mathcal C$, whose transitions are labelled $1$,
$e^+$ or $e^-$ for transitions $e$ of $M$, with the following property. For
complex numbers $\beta_e$, give these labels the weights $1$, $\beta_e$ and
$\overline{\beta_e}$. On input $(x,T)$, with the advice of $M$ on its advice
tape (and, in the Merlin--Arthur case, $u,v\in\bits^w$, $w\le T$, on the
input), every run of $M'$ halts within $T'=O_M(T)$ steps and makes exactly
$c_\Delta T$ binary choices, and the weighted count of its accepting choice strings
in $\bits^{T'}$ is
\[
 A^\beta_{\mathrm{acc}}(x,T)=2^{T'-c_\Delta T}\sum_{c}\ \sum_{\substack{p_1,p_2:c_0\to c\\
 m(p_1(t))=m(p_2(t))\ \forall t\le T}}\ \prod_{t=1}^T
 \beta_{e_t(p_1)}\overline{\beta_{e_t(p_2)}},
\]
with $c$ ranging over the configurations with control state in $\mathcal C$
(respectively, with $p_1$ starting from witness $v$ and $p_2$ from witness
$u$). With $\beta_e=D\alpha_e$ the sum is $Z_{\mathcal C}(x,T)$
(respectively $Z(u,v)$).
\end{lemma}

\begin{proof}
The machine $M'$ keeps two copies of $M$, the \emph{ket} copy on work tapes
$1,\dots,k$ and the \emph{bra} copy on tapes $k+1,\dots,2k$. It first copies
the first $T+1$ symbols of the input and of the advice onto separate work
tapes for each copy, so that each copy has its own read-only heads, and
writes $v$ onto the ket copy of tape $1$ and $u$ onto the bra copy; this is
deterministic and takes $O(T)$ steps. It then runs $T$ rounds. In a round,
for each copy it chooses a transition applicable to that copy's local
situation, encoded in exactly $c_\Delta/2$ binary choices with
$c_\Delta=2\lceil\log_2(\Delta+1)\rceil$; a code that names no applicable
transition sets a reject flag. A halted copy has only its stationary
transition. The step that applies $e$ to the ket copy has label $e^+$, the
step that applies $e$ to the bra copy has label $e^-$, and all other steps
have label $1$. The machine sets the reject flag if the two copies give
different values of the measured function $m$. After $T$ rounds, $M'$ checks
that the two copies are in the same configuration with control state in
$\mathcal C$: same control state, same head positions, same tape contents.
Each copy of a tape carries a track that marks its starting cell, its
visited cells and its head cell. For each tape, $M'$ moves both copies to
the starting mark, then sweeps both to the left and then to the right in
lockstep until both are past their visited cells, comparing symbols and head
marks cell by cell; this takes $O(T)$ steps. It accepts iff the reject
flag is off and the configurations agree.

Every run makes exactly $c_\Delta T$ choices, and its other steps ignore their
choice bits, which gives the factor $2^{T'-c_\Delta T}$. An accepting choice
pattern determines one pair of paths $(p_1,p_2)$ with equal measured values
ending in a common configuration in $\mathcal C$, every such pair arises from
exactly one pattern, and the weight of the run is
$\prod_t\beta_{e_t(p_1)}\overline{\beta_{e_t(p_2)}}$.
\end{proof}

The weighted count is computed by the contraction over a finite ring, and the
sign of the result is read off from its coordinates.

\begin{lemma}\label{lem:sign}
Let $\mathbb K\subset\C$ be a fixed number field with integral basis
$\gamma_1,\dots,\gamma_d$. Let $Y=\sum_jy_j\gamma_j\in\mathcal O_{\mathbb
K}\cap\mathbb R$ with $|y_j|<2^M$ for a known $M\le\poly(T)$, and suppose that
the residue of $Y$ in $\mathcal O_{\mathbb K}/p\mathcal O_{\mathbb K}$ can be
computed in space $S\ge\log(T+2)$ for every admissible odd prime $p$ of
$O(\log(T+2))$ bits. Then every bit of every $y_j$, and the sign of $Y$, can
be computed in space $O_{\mathbb K}(S+\log(T+2))$.
\end{lemma}

\begin{proof}
The residue of $Y$ in the basis gives the residues of the $y_j$. The first
$2M+10$ admissible odd primes have $O(\log(T+2))$ bits and product above
$2^{2M+5}$, so Chinese remaindering as in Section~\ref{sec:reconstruction},
applied to $y_j+2^M$, gives any bit of $y_j$. If all $y_j$ vanish, $Y=0$.
Otherwise let $\sigma_1,\dots,\sigma_d$ be the embeddings of $\mathbb K$,
with $\sigma_1$ the given one, and $c=\max_{i,j}|\sigma_i(\gamma_j)|$. The norm
$\prod_i\sigma_i(Y)$ is a nonzero rational integer and
$|\sigma_i(Y)|\le dc2^M$, so $|Y|\ge(dc2^M)^{-(d-1)}$. Since $Y$ is real,
$Y=\sum_jy_j\operatorname{Re}\gamma_j$. Let $q=d(M+\lceil\log(dc)\rceil+2)$
and let $r_j$ be dyadic numbers with $|r_j-\operatorname{Re}\gamma_j|\le
2^{-q}$. Then $|\sum_jy_jr_j-Y|\le d2^{M-q}<|Y|/2$, so $Y$ and $\sum_jy_jr_j$
have the same sign. Each $\operatorname{Re}\gamma_j$ is a fixed real
algebraic number, so its first $q$ bits are computed from $1^q$ in uniform
$\mathrm{TC}^0$ by Je\v{r}\'abek's root approximation~\cite{Jer12} (the
binary expansion of an algebraic constant, with the position given in
unary), hence in space $O(\log q)$.
Iterated multiplication and addition~\cite{HAB02}, applied to these bits and
to the bits of the $y_j$ generated on demand, give the sign of
$\sum_jy_jr_j$ with $O(\log(T+2))$ further bits.
\end{proof}

\begin{remark}
For $\mathbb K=\mathbb Q(\zeta_8)$ no root approximation is needed. Then
$Y\in\Z[\zeta_8]\cap\mathbb R=\Z[\sqrt2]$, and in the basis
$1,\zeta_8,\zeta_8^2,\zeta_8^3$ one has $Y=A+B\zeta_8-B\zeta_8^3=A+B\sqrt2$.
If $A,B\ge0$ then $Y\ge0$, and if $A,B<0$ then $Y<0$. If $A\ge0>B$, then
$Y\ge0$ iff $A^2\ge2B^2$; if $B\ge0>A$, then $Y\ge0$ iff $A^2\le2B^2$.
Chinese remaindering gives the signs of $A$, $B$ and $A^2-2B^2$.
\end{remark}

\begin{proof}[Proof of Theorem~\ref{thm:quantum-main}(1)] Take
$R=\mathcal O_{\mathbb K}$ and $\beta_e=D\alpha_e$, a fixed table whose
residues modulo $p$ take $O(\log(T+2))$ bits. For $\mathcal C$ the accepting,
the postselecting, or the nonhalting control states,
Lemma~\ref{lem:quantum-paths} and Section~\ref{sec:reconstruction} give the
residue of $2^{T'-c_\Delta T}Z_{\mathcal C}$ in $R_p$, hence that of $Z_{\mathcal C}$,
for every admissible odd prime, in space $O_M(T/\sqrt{\log(T+2)})$; the
constants depend only on $M$, because $M'$ does. For coordinate bounds use
the $\ell_1$ norm $\|\cdot\|_1$ of coordinates in the integral basis. The
structure constants of $\mathcal O_{\mathbb K}$ give
$\|yz\|_1\le c_{\mathbb K}\|y\|_1\|z\|_1$, every term of $Z_{\mathcal C}$ is a
product of $2T$ table entries, and there are at most $\Delta^{2T}$ pairs of
paths, so all coordinates are below $2^M$ with $M=O_M(T)$. The quantities
$rZ_{\mathrm{acc}}-jD^{2T}$ and $2Z_{\mathrm{acc}}-Z_{\mathrm{post}}$ are
real algebraic integers with coordinates below $2^{M+O(\log T)}$; the
residues of the integer $D^{2T}$ follow by repeated squaring. So
Lemma~\ref{lem:sign} gives their signs and the bits of all coordinates.
Bounded error is the test $2Z_{\mathrm{acc}}\ge D^{2T}$ and postselection the
test $2Z_{\mathrm{acc}}\ge Z_{\mathrm{post}}$. Without a supplied clock,
double $T$ until $Z_{\mathrm{nonhalt}}=0$, that is, until all of its
coordinates vanish; the peak cap is at most $2\max\{2,t_M(x)\}$.
\end{proof}

\begin{proof}[Proof of Theorem~\ref{thm:quantum-main}(2)] Fix $T$. Every
column of $V$ has at most $\Delta$ nonzero entries. So does every row, up to
a constant $s_M\ge\Delta$: a configuration has $O_M(1)$ predecessors, since
the old head positions are within one cell of the new ones, the old control
state and the old scanned symbols range over a finite set, and all other
cells are unchanged. Let $h=\lceil\log_2(T+2)\rceil+c_M$ for a constant $c_M$
chosen below. For each nonstationary transition compute a Gaussian dyadic
$\widetilde\alpha_e$ with $2^h\widetilde\alpha_e\in\Z[\mathrm i]$ and
$|\widetilde\alpha_e-\alpha_e|\le2^{-h}$; this takes time and space
$\poly(\log T)$, and the $O_M(1)$ results are stored in $O_M(\log(T+2))$
bits. Stationary transitions keep amplitude $1$. Let $\widetilde V$ be the
operator built from the $\widetilde\alpha_e$. In each column of
$\widetilde V-V$ the absolute values add up to at most $\Delta2^{-h}$, since
at most $\Delta$ transitions leave a configuration, and so in each row they
add up to at most $s_M\Delta2^{-h}$. Hence
\[
 \|\widetilde V-V\|\le\sqrt{\|\widetilde V-V\|_{1\to1}
 \|\widetilde V-V\|_{\infty\to\infty}}\le s_M\Delta2^{-h}=:\delta,
\]
and we choose $c_M$ so that $\delta\le1/(200T)$.

Put $\Phi(X)=\sum_\mu V\Pi_\mu X\Pi_\mu V^\dagger$ and define
$\widetilde\Phi$ with $\widetilde V$. The pinching $X\mapsto\sum_\mu\Pi_\mu
X\Pi_\mu$ does not increase the trace norm, $\|V\Pi_\mu\|\le1$ and
$\|\widetilde V\Pi_\mu-V\Pi_\mu\|\le\delta$. Hence $\|\Phi(X)\|_1\le\|X\|_1$,
$\|\widetilde\Phi(X)\|_1\le(1+\delta)^2\|X\|_1$ and
$\|(\widetilde\Phi-\Phi)(X)\|_1\le\delta(2+\delta)\|X\|_1$ for every matrix
$X$. Writing $\widetilde\Phi^T-\Phi^T=\sum_{k<T}\widetilde\Phi^{T-1-k}
(\widetilde\Phi-\Phi)\Phi^k$ gives, for $\rho_0=|c_0\rangle\langle c_0|$,
\[
 \|\widetilde\Phi^T(\rho_0)-\Phi^T(\rho_0)\|_1\le
 T\delta(2+\delta)(1+\delta)^{2T}<\tfrac1{48}.
\]
So for every set $\mathcal C$ the number
$\widetilde p_{\mathcal C}=\operatorname{tr}(\Pi_{\mathcal C}
\widetilde\Phi^T(\rho_0))$ is within $\frac1{48}$ of the true probability
$p_{\mathcal C}$ of being in $\mathcal C$ at step $T$. Expanding
$\widetilde\Phi^T$ along basis configurations shows that
$\widetilde Z_{\mathcal C}=2^{2hT}\widetilde p_{\mathcal C}$ is the sum of
Lemma~\ref{lem:quantum-paths} with $\beta_e=2^h\widetilde\alpha_e$ (and
$\beta=2^h$ for stationary transitions). It is real and lies in
$\Z[\mathrm i]$, so it is an integer, of absolute value at most
$\Delta^{2T}2^{2T(h+1)}=2^{O_M(T\log T)}$. Run the contraction over
$R_p=\F_p[y]/(y^2+1)$, which is $\Z[\mathrm i]/p$, and recover any bit of
$\widetilde Z_{\mathcal C}$ by Chinese remaindering with $O_M(T\log T)$
admissible primes; Section~\ref{sec:reconstruction} allows polynomially many.
The approximate operator need not be unitary, and nothing above uses that it
is. Rounding happens once per table entry, not once per path.

With a supplied cap, accept iff $2\widetilde Z_{\mathrm{acc}}\ge2^{2hT}$;
under the promise $p_{\mathrm{acc}}\notin(\frac13,\frac23)$ this is correct
because $|\widetilde p_{\mathrm{acc}}-p_{\mathrm{acc}}|<\frac1{48}$. Without a
cap, double $T$ from $2$ and stop at the first $T$ with
$16\widetilde Z_{\mathrm{nonhalt}}\le2^{2hT}$, then accept iff
$2\widetilde Z_{\mathrm{acc}}\ge2^{2hT}$. Let $p$ be the probability that
$M$ eventually accepts. Halted branches keep their outcome, so the
probability of accepting by step $T$ lies in $[p-p_{\mathrm{nonhalt}},p]$. At
the stopping cap $p_{\mathrm{nonhalt}}\le\frac1{16}+\frac1{48}=\frac1{12}$.
If $p\ge\frac23$ then $\widetilde p_{\mathrm{acc}}\ge\frac23-\frac1{12}-
\frac1{48}=\frac{27}{48}$, and if $p\le\frac13$ then $\widetilde
p_{\mathrm{acc}}\le\frac13+\frac1{48}=\frac{17}{48}$, so the answer is
correct. If $T$ is at least the worst-case time $t$, then
$p_{\mathrm{nonhalt}}=0$ and $\widetilde p_{\mathrm{nonhalt}}<\frac1{48}$, so
the loop stops by $T<2\max\{2,t\}$. If $t$ is the expected halting time,
Markov's inequality gives $p_{\mathrm{nonhalt}}\le t/T\le\frac1{24}$ for
$T\ge24t$, so $\widetilde p_{\mathrm{nonhalt}}<\frac1{16}$ and the loop stops
by $T<48\max\{2,t\}$. Each cap reuses the space of the previous one.
\end{proof}

\begin{remark}
The same stopping rule, with the exact tests $16A_{\mathrm{nonhalt}}\le2^T$
and $16Z_{\mathrm{nonhalt}}\le D^{2T}$, extends Theorem~\ref{thm:main} and
Theorem~\ref{thm:quantum-main}(1) to bounded-error machines whose
\emph{expected} running time on $x$ is $t$, in space
$O_M(t/\sqrt{\log(t+2)})$.
\end{remark}

\begin{proof}[Proof of Corollary~\ref{cor:qma}] Let $\widetilde E=D^{2T}E_x$, a
Hermitian matrix with entries $Z(u,v)\in\mathcal O_{\mathbb K}$, and let $k$
be the even integer in $\{w+1,w+2\}$. Let $\lambda_{\max}$ be the largest
eigenvalue of $E_x$. Then
$\lambda_{\max}^k\le\operatorname{Tr}(E_x^k)\le2^w\lambda_{\max}^k$, so
$\operatorname{Tr}(E_x^k)\ge(2/3)^k$ if the verifier accepts some witness
with probability $\frac23$, and $\operatorname{Tr}(E_x^k)\le2^w3^{-k}\le
\frac12(2/3)^k$ if it accepts every witness with probability at most
$\frac13$. So the verifier is simulated by the test
$4\cdot3^k\operatorname{Tr}(\widetilde E^k)\ge3\cdot2^kD^{2kT}$. Trace-power
tests of this kind are standard; we only check the space.

By Lemma~\ref{lem:quantum-paths} with $\beta_e=D\alpha_e$, the residue of
$Z(u,v)$ in $R_p=\mathcal O_{\mathbb K}/p\mathcal O_{\mathbb K}$ is computable
in space $O(T/\sqrt{\log(T+2)})$, with $u,v$ read from the caller. Compute
the residue of $\operatorname{Tr}(\widetilde E^k)$ as
$\sum_u(\widetilde E^k)_{u,u}$, evaluating $(\widetilde E^r)_{a,b}=
\sum_c(\widetilde E^{\lfloor r/2\rfloor})_{a,c}
(\widetilde E^{\lceil r/2\rceil})_{c,b}$ recursively. The recursion has depth
$O(\log k)$, each frame holds three strings in $\bits^w$, an exponent and
$O(\log p)$-bit accumulators, and the base case is one entry of $\widetilde
E$; this gives the stated space. The trace lies in $\mathcal O_{\mathbb
K}\cap\mathbb R$. As in the proof of Theorem~\ref{thm:quantum-main}(1),
$\|Z(u,v)\|_1<2^{c_1T}$ for a constant $c_1$, so the entries of
$\widetilde E^k$ have norm below $(c_{\mathbb K}2^w)^{k-1}2^{c_1Tk}$ and the
coordinates of the trace are below $2^{M_k}$ with
$M_k=O_{\mathbb K}(w+k(w+c_1T))=\poly(T)$. Lemma~\ref{lem:sign} decides the
test.
\end{proof}

\begin{remark}
With a classical witness of $w$ bits, or with a constant number of
alternating classical quantifiers over strings of total length $w$,
enumerating the strings outside the simulation of
Theorem~\ref{thm:quantum-main} gives space $O(T/\sqrt{\log(T+2)}+w)$.
\end{remark}

\section{Circuits}
\label{sec:circuits}

A circuit of $S$ gates over a fixed gate set is a list of gates; each gate
names its qubits (or, for Boolean circuits, its input wires) by indices, and
$L$ is the length of the whole description, including the classical input.
For a quantum circuit, let $\mathbb K\subset\C$ be a number field closed
under complex conjugation that contains all gate entries, and $D\ge1$ an
integer with $D$ times every entry in $\mathcal O_{\mathbb K}$. Each qubit
has a prescribed initial basis value, read from the input or set to $0$; in
the Merlin--Arthur case $w$ designated qubits instead hold a witness. A
measurement gate measures one qubit in the computational basis. At the end
the designated qubits are measured, and $\tau$ is a prescribed assignment to
them; acceptance, postselection, and acceptance with postselection are such
events. Write
\[
 Z_\tau=D^{2S}\Pr[\text{the designated qubits read }\tau],
\]
and $Z_\tau(u,v)=D^{2S}\langle u|E_\tau|v\rangle$ in the Merlin--Arthur case.

\begin{lemma}\label{lem:circuit-quantum}
Let $C$ be a quantum circuit of $S\ge2$ gates as above, whose gates are
generated in space $s\ge\log(L+2)$, and let $1\le g\le S$. Either
$Z_\tau(u,v)=0$, which is detected in space $O(s)$, or for every admissible
odd prime $p$ the residue of $Z_\tau(u,v)$ in $\mathcal O_{\mathbb
K}/p\mathcal O_{\mathbb K}$ is the value of one network on $g$ vertices with
$P=8\lceil S/g\rceil$ and $U\le gP=O(S)$, whose port lists are generated in
auxiliary space $O(s)$ and whose vertex weights are evaluated in auxiliary
space $O(\lceil S/g\rceil+s)$.
\end{lemma}

\begin{proof}
For each witness or designated qubit, scan the gate list to see whether a
gate touches it; other untouched qubits contribute a factor $1$. Such a qubit keeps
its initial value in both the ket and the bra copy. It contributes a factor
$1$, unless it is a witness qubit with $u_q\ne v_q$ or a designated qubit
whose initial value differs from $\tau_q$; in those cases $Z_\tau(u,v)=0$.

Split the gates $1,\dots,S$ into $g$ contiguous blocks of at most
$b_0=\lceil S/g\rceil$ gates, and put both the ket and the bra copy of gate
$i$ into vertex $a=\lceil i/b_0\rceil$. The setting $z_a$ lists, for each gate
of the block, the ket and the bra values of its input and output qubits, at
most $8$ bits per gate. For each touched qubit $q$, let $i_1<\dots<i_r$ be
the gates acting on $q$. At $i_1$, which is recognized by scanning the
earlier gates, the ket and bra input values of $q$ are its initial values
($v_q$ and $u_q$ for a witness qubit). At $i_r$ the ket and bra output
values of $q$ are equal, which is the partial trace, and equal $\tau_q$ if
$q$ is designated; neither condition needs an edge. For consecutive gates
$i_m,i_{m+1}$ in the same block the vertex checks equality of the output
values at $i_m$ with the input values at $i_{m+1}$; in different blocks this
is a $2$-bit equality edge. A gate has at most two incoming and two outgoing
such edges, so the port width of a vertex is at most $8b_0=P$ and
$U\le gP\le8(S+g)$.

A unitary gate with matrix $G$ contributes
$DG_{y,x}\cdot\overline{DG_{y',x'}}$ for ket values $x\to y$ and bra values
$x'\to y'$. A measurement gate requires ket input, ket output, bra input and
bra output to be equal and contributes $D^2$, so that every gate carries the
factor $D^2$. A vertex weight is the product of these factors over its gates,
or $0$ if a local check fails, and is evaluated in space $O(b_0+s)$. Scanning
the block and finding the next gate on each qubit lists the ports in space
$O(s)$. Expanding the network gives exactly the doubled path sum for
$Z_\tau(u,v)$.
\end{proof}

\begin{lemma}\label{lem:circuit-boolean}
Let $C(x,y)$ be a fan-in-$2$ Boolean circuit of $S\ge2$ gates with
unrestricted fan-out, output gate $S$, and gates generated in space
$s\ge\log(L+2)$, and let $x$ be fixed. Let $n_0$ be the number of variables
of $y$ read by no gate. For $1\le g\le S$, the number
$2^{-n_0}|\{y:C(x,y)=1\}|$ is the value of one network on $g$ vertices with
Boolean weights, $P=5\lceil S/g\rceil$ and $U\le gP=O(S)$, whose port lists
are generated in auxiliary space $O(s)$ and whose vertex predicates are
evaluated in auxiliary space $O(\lceil S/g\rceil+s)$.
\end{lemma}

\begin{proof}
For each wire, meaning a variable or a gate output, let $i_1<\dots<i_r$ be
the distinct gates that read it. Chain the wire through its readers: a gate
output passes from its gate to $i_1$, then from $i_1$ to $i_2$, and so on; a
variable starts at $i_1$, where it is free if it is a variable of $y$ and is
checked against the input if it is a variable of $x$. A gate reading the
same wire twice checks that its two inputs agree. Each gate holds $3$ setting
bits, its two inputs and its output, checks them against its truth table, and
at gate $S$ requires output $1$. It receives at most $2$ chain bits and sends
at most $3$: its output to the first reader, and each of its inputs to the
next reader of that wire. No copy trees are needed, so every gate has port
width at most $5$. Contiguous blocks of $b_0=\lceil S/g\rceil$ gates give
$P=5b_0$ and $U\le5(S+g)$. An assignment to the variables of $y$ that are
read determines exactly one passing tuple of settings, including the gates
whose output nobody reads.
\end{proof}

\begin{proof}[Proof of Theorem~\ref{thm:circuits}] Run the pipeline of
Section~\ref{sec:reconstruction} with $S$ in place of $T$: choose $j$ maximal
with $K_j\le S^{1/D_1}$, put $g=3K_j^8$ (for small $S$ enumerate
exhaustively), and feed the network of Lemma~\ref{lem:circuit-quantum} or
Lemma~\ref{lem:circuit-boolean} to Lemma~\ref{lem:group} (first case,
$\zeta=1/(2K_j)$) and Lemma~\ref{lem:clique}. That grouping case places no
bound on the individual pair widths $c_{ab}$, so it applies to arbitrary
wiring. There is only one network, and gate lookup costs $O(s)$ more. For
part~(1), the network has at most $2^{8S}$ settings and each term is a
product of $S$ factors from a fixed finite set, so the coordinates of
$Z_\tau$ are below $2^{O(S)}$, and Lemma~\ref{lem:sign} gives their bits and
the signs of $rZ_\tau-jD^{2S}$ and $rZ_\tau-jZ_{\tau'}$. For
part~(2), at most $2S$ variables are read, so the count in
Lemma~\ref{lem:circuit-boolean} is at most $2^{2S}$, and Chinese remaindering
with $2S+2$ primes gives its bits; the bits of $|\{y:C(x,y)=1\}|$ are those
bits shifted by $n_0$.
\end{proof}

\begin{remark}
\emph{Time.} A halting machine in space $S'\ge\log L$ on an input of
length $L$ runs in time $L\,2^{O(S')}$, so Theorem~\ref{thm:circuits} gives time
$2^{O(S/\sqrt{\log S})}\poly(L)$. For exact amplitudes of Clifford$+T$
circuits this time bound is due to Tulsyan~\cite{Tul26}, who derives it
from~\cite{Cla26}; his main theorem states no space bound. For uniform
families over Hadamard and Toffoli gates, the time bound with polynomial
space is also in~\cite{CSL26}. As far as we know the space bound is new. \emph{One
amplitude.} An amplitude $\langle y|C|x\rangle$ needs one copy of the wire
network, with both ends fixed, instead of the doubled one.
\emph{Equivalence.} For circuits $C_1,C_2$ on $n$ qubits, pairing every
touched qubit with a fresh partner in a Bell pair gives a circuit of $O(S)$
gates whose acceptance probability is
$|\operatorname{Tr}(C_1^\dagger C_2)|^2/2^{2n}$; untouched qubits contribute
a factor $1$. Part~(1) then decides equivalence up to a global phase and
every rational fidelity threshold, after enlarging $\mathbb K$ to contain
$\sqrt2$ and closing the gate set under adjoints.
\emph{Merlin--Arthur.} The proof of Corollary~\ref{cor:qma}, with
Lemma~\ref{lem:circuit-quantum} at the leaves, decides circuit verifiers with
a $w$-qubit witness in space
$O(S/\sqrt{\log(S+2)}+(w+\log(S+2))\log(w+2)+\log(L+2))$.
\emph{Uniform families.} If gate $i$ of the circuit for length $n$ is
computed from $(1^n,i)$ in space $O(\log(n+S))$, the bounds hold with
$\log(n+S)$ in place of $\log(L+2)$, and advice can be read as further
classical input. So languages decided with bounded error or with
postselection by logspace-uniform quantum circuits of size $S(n)$, with $a$
bits of advice, are in $\DSPACE(S/\sqrt{\log(S+2)}+\log n)/a$.
\end{remark}

\section{Separations from linear space}
\label{sec:separations}

\begin{theorem}\label{thm:separations}
Unconditionally:
\begin{enumerate}
\item For every constant $K$ there is a language $L_K\in\DSPACE(n)$ such that
      no language in $\PostBQTIME[t]/(Kn)$, $\BQTIME[t]/(Kn)$,
      $\PostBPTIME[t]/(Kn)$ or $\BPTIME[t]/(Kn)$ with $t=o(n\sqrt{\log n})$,
      and no language decided with bounded error or with postselection by
      logspace-uniform quantum circuits of size $o(n\sqrt{\log n})$ over a
      fixed algebraic gate set with $Kn$ bits of advice, agrees with $L_K$
      on infinitely many input lengths.
\item There is a language $L^*\in\DSPACE(n)$ such that for every quantum
      (with algebraic amplitudes) or randomized multitape machine $M$ of
      worst-case time $o(n\sqrt{\log n})$, and every logspace-uniform
      quantum circuit family $M$ of size $o(n\sqrt{\log n})$ over a fixed
      algebraic gate set, with no bounded-error promise, and every advice
      sequence $w_n$ of
      length at most $\log^2n$, for all large $n$,
      \[
       \Bigl|\Pr_{x\sim U_n}\bigl[M(x;w_n)=L^*(x)\bigr]-\tfrac12\Bigr|
       \le O\Bigl(\frac{\log n}{\sqrt n}\Bigr).
      \]
\end{enumerate}
\end{theorem}

\begin{proof}
Both languages diagonalize against binary-alphabet deterministic machines
$M_1,M_2,\dots$ with read-only inputs, run by a universal interpreter with a
work-space cap $S(n)=\lfloor n/4\rfloor$ and a step counter of
$S(n)+3\log n$ bits, outputting $0$ on overflow. Put
$f(n)=\lfloor\log\log n\rfloor$. A fixed machine with $k$ work tapes, of
space $s(n)=o(n)$, that halts on an input of length $n$ with advice of
length at most $Kn$ (or $\log^2n$) runs for at most
$|Q|(n+2)(Kn+2)\,s(n)^k2^{ks(n)}$ steps, since it never repeats a
configuration; for large $n$ this is below $2^{S(n)}n^3$, and $ks(n)\le S(n)$.
So such a machine halts within these caps for all large $n$, and it has
index at most $f(n)$.

\emph{(1).} Put $m=Kn+\lceil\log f(n)\rceil+2$ and let $x_1,\dots,x_m$ be the
first $m$ strings of length $n$. Let $v^*\in\bits^m$ be the
lexicographically first vector such that for no $i\le f(n)$ and no
$w\in\bits^{\le Kn}$ does $M_i(x_j;w)=v^*_j$ hold for all $j\le m$; it exists
because $f(n)(2^{Kn+1}-1)<2^m$. Put $L_K(x_j)=v^*_j$ and $L_K=0$ elsewhere.
Finding $v^*$ takes $O_K(n)$ bits, so $L_K\in\DSPACE(n)$ by linear
compression. By Theorems~\ref{thm:main}, \ref{thm:quantum-main}
and~\ref{thm:circuits} (with the remark on uniform families) every language
in these classes is in $\DSPACE(o(n))/(Kn)$; here
$t/\sqrt{\log(t+2)}=o(n)$, using $\log t\ge\log n$ for $t\ge n$ and the
monotonicity of $t/\sqrt{\log(t+2)}$ for $t<n$, and the same holds for the
circuit size in place of $t$. A fixed machine of space
$o(n)$ with advice $w_n$ is some $M_i$ with $w=w_n$ at every large length,
so it disagrees with $L_K$ on $x_1,\dots,x_m$.

\emph{(2).} Put $\mu=\lfloor\log(n/4)\rfloor$, $\mu'=2^\mu\in(n/8,n/4]$,
$K(n)=\lceil\sqrt n\rceil$, $\ell(n)=\lfloor\log^2n\rfloor$ and
$\Delta(n)=\lceil\sqrt{\mu'(\ell(n)+\lceil\log(f(n)K(n))\rceil+4)}\rceil$.
Write an input as $x=(y,z)$ with $y\in\bits^\mu$. For $z\in\bits^{n-\mu}$,
let $v^*(z)$ be the lexicographically first $v\in\bits^{\mu'}$ such that for
every $i\le f(n)$, $j\in\{1,\dots,K(n)-1\}$ and $w\in\bits^{\le\ell(n)}$, the
vector $u=(M_i((y',z),j,w))_{y'\in\bits^\mu}$ agrees with $v$ in
$\mu'/2\pm\Delta(n)$ positions (strictly inside). By Hoeffding's inequality a
uniform $v$ fails a given triple with probability less than
$2^{-(\ell+\lceil\log(fK)\rceil+3)}$, and there are fewer than $2fK2^\ell$
triples, so $v^*(z)$ exists. Put $L^*(y,z)=v^*(z)_y$. Reading $z$ from the
input, this takes $\mu'+2S(n)+O(\log^2n)\le\frac34n+o(n)$ bits, so
$L^*\in\DSPACE(n)$ even over the binary alphabet.

Let $M$ and $w_n$ be as in~(2), with time bound $t=o(n\sqrt{\log n})$. For
$\beta\in\bits$ let $A_\beta(x,j,w)$ output $1$ iff
$\Pr[M(x;w)\oplus\beta=1]\ge j/K(|x|)$. By Theorems~\ref{thm:main},
\ref{thm:quantum-main}(1) and~\ref{thm:circuits} this exact threshold test,
with the cap found by doubling for machines, runs in space $o(n)$ and
halts when the advice is $w_n$, so for large $n$ both $A_\beta$ are among the
machines diagonalized against. (On other advice strings the doubling may not
stop, and the caps of the interpreter cut the run off.) Hence for every $j$,
$\Pr_{x\sim U_n}[A_\beta(x,j,w_n)=L^*(x)]<\frac12+\Delta(n)/\mu'$. For
$q\in[0,1]$ and $b\in\bits$,
\[
 [b=1]\,q+[b=0]\,(1-q)\le\frac1K+\frac1K\sum_{j=1}^{K-1}
 \bigl[[q\ge j/K]=b\bigr],
\]
because at least $Kq-1$ of the indices $j$ satisfy $j\le Kq$ and at least
$K(1-q)-1$ satisfy $j>Kq$. Apply this with $q=\Pr[M(x;w_n)\oplus\beta=1]$ and
$b=L^*(x)$, and average over $x$:
$\Pr[M(x;w_n)\oplus\beta=L^*(x)]<\frac12+\frac1{K(n)}+\frac{\Delta(n)}{\mu'}
=\frac12+O((\log n)/\sqrt n)$. The cases $\beta=0$ and $\beta=1$ give the two
sides of~(2).
\end{proof}

Taking $\ell(n)=\lfloor n^{1-2c}\rfloor$ in~(2) gives correlation
$O(n^{-c})$ against $n^{1-2c}$ bits of advice, for $0<c<\frac12$.

Pointwise tape-dependent advice~\cite{TV07} lets the advice depend on the
random tape: $L\in\BPTIME[t]//_{\mathrm{pt}}a$ if a deterministic time-$t$
machine $M(x;R,w)$ and maps $\alpha_n:\bits^{t}\to\bits^{\le a(n)}$ satisfy
$\Pr_R[M(x;R,\alpha_n(R))\neq L(x)]\le\frac13$ for every $x\in\bits^n$.
Advice of this kind cannot be removed by trying all advice strings.

\begin{corollary}\label{cor:pt-advice}
With unrestricted coins, $\DSPACE(n)\not\subseteq\allowbreak\BPTIME[o(n\sqrt{\log n})]
//_{\mathrm{pt}}O(1)$, and for every $\varepsilon>0$,
$\DSPACE(n)\not\subseteq\allowbreak\BPTIME[O(n)]//_{\mathrm{pt}}
((\frac12-\varepsilon)\log\log n)$.
\end{corollary}

\begin{proof}
Fix a machine $M(x;R,w)$, a time cap $T_0$ and an advice cap $\ell$, put
$L=162(\ell+6)$, and let $x_1,\dots,x_L$ be the first $L$ strings of length
$n$. For $v\in\bits^L$ let $u(R,w)=(M(x_i;R,w))_{i\le L}$,
\[
 D_v(R)=\Bigl[\exists w\in\bits^{\le\ell}:\
 \mathrm{dist}_H(u(R,w),v)\le\lfloor4L/9\rfloor\Bigr],\qquad
 p_v=\Pr_{R\in\bits^{T_0}}[D_v(R)=1].
\]
If some advice map makes $M$ compute $v$ on $x_1,\dots,x_L$ with pointwise
error $\frac13$, the expected distance is at most $L/3$, so $p_v\ge\frac14$
by Markov's inequality. On the other hand, a Hamming ball of radius
$\lfloor4L/9\rfloor$ has at most $2^{(1-1/162)L}$ points, since
$H_2(\frac12-\gamma)\le1-2\gamma^2$; for each $R$ fewer than $2^{\ell+1}$
balls cover the $v$ with $D_v(R)=1$, so
$2^{-L}\sum_vp_v<2^{\ell+1-L/162}=\frac1{32}$ and some $v$ has
$p_v<\frac1{16}$.

The machine $\widetilde M_v$ that writes $T_0$ coins on a work tape and then
deterministically evaluates $D_v$, running $M$ on each $(x_i,w)$ with the
stored coins, takes time $T_{\mathrm{test}}=O_M(2^\ell LT_0)$ and accepts with
probability exactly $p_v$. It reads $v$, $\ell$ and $T_0$ from its advice
tape, so its control does not depend on them, and by
Theorem~\ref{thm:main} the test $p_v<\frac1{16}$ takes space at most
$C_M2^\ell LT_0/\sqrt{\log(T_0+2)}$ with $C_M$ depending only on $M$. The
diagonal language takes, on a length assigned to $(M,\ell,T_0)$, the values
of the first $v$ with $p_v<\frac1{16}$ on $x_1,\dots,x_L$ and $0$ elsewhere,
and outputs $0$ if its work space would exceed $n$ cells. It runs the
algorithm of Theorem~\ref{thm:main} for $\widetilde M_v$ directly from the
description of $M$, so the constant is $C_M$ and no simulation of $M$ by a
machine with a fixed number of tapes is involved.

For the first claim, assign infinitely many lengths to every triple
$(i,c,r)$ of positive integers, with $M=M_i$, $\ell=c$ and
$T_0=\lfloor n\sqrt{\log(n+2)}/r\rfloor$. A machine of time
$o(n\sqrt{\log n})$ with $c$ bits of advice finishes within $T_0$ for large
$n$, and for $r$ large compared with $C_{M_i}2^cL$ the test fits in $n$
cells, so the language defeats it. For the second claim assign lengths to
pairs $(i,c)$ with $T_0=cn$ and $\ell=\lfloor(\frac12-\varepsilon)\log\log
n\rfloor$; the space is $O_{i,c}(n(\log n)^{-\varepsilon}\log\log n)=o(n)$.
\end{proof}

The gain over linear time is a $\sqrt{\log n}$ factor. At
$t=n^{2-\varepsilon}$ with $0<\varepsilon<1$ the space bounds of
Theorems~\ref{thm:main} and~\ref{thm:quantum-main} still exceed $n$, so these
results say nothing about $\DSPACE(n)$ against $\BPTIME(n^{1+\varepsilon})$.

\section{\texorpdfstring{Exact counting stops at space $T^{1-o(1)}$}{Exact counting stops at space T\^{}(1-o(1))}}
\label{sec:limits}

Theorems~\ref{thm:main} and~\ref{thm:quantum-main} compute exact counts, and
in particular decide whether $A(x,T)>0$. The following observation shows
that neither exact counting nor bounded-error postselected simulation can
save a polynomial factor in space unless a hypothesis that survived the
refutation of ETH fails. The \emph{Weak Exponential Time Hypothesis} states
that for every constant $\varepsilon>0$, 3SAT on $n$ variables has no
$O(2^{n^{1-\varepsilon}})L^{O(1)}$-time algorithm, where $L$ is the input
length. The MIT Hardness Group~\cite{MIT26} proved it equivalent to the
Stearns--Hunt Satisfiability Hypothesis and showed that it is preserved by
reductions of polynomial blowup; with the identity reduction (blowup
exponent~$1$), it rules out $O(2^{s^{1-\varepsilon}})$-time algorithms for
3SAT formulas of size $s=\Theta(n+m)$, for every constant $\varepsilon>0$.

\begin{proposition}\label{prop:limits}
Suppose there is a constant $\varepsilon>0$ such that, for $T\ge|x|$, either
\begin{enumerate}
\item for every multitape machine $M$ with at most one binary choice per
      step, whether $A(x,T)>0$ can be decided in deterministic space
      $O(T^{1-\varepsilon})$, or
\item the class $\PostBPTIME[T]$ is contained in
      $\DSPACE(O(T^{1-\varepsilon}))$, as would follow from the same
      containment for $\PostBQTIME[T]$, or
\item for explicit fan-in-$2$ Boolean circuits of $S$ gates and description
      length $L$, satisfiability, or bounded-error postselected acceptance
      with the inputs as coins, can be decided in deterministic space
      $O(S^{1-\varepsilon}+\log L)$.
\end{enumerate}
Then the Weak Exponential Time Hypothesis fails.
\end{proposition}

\begin{proof}
A 3SAT formula of size $s$ has an encoding of length $L=O(s\log s)$. A
multitape machine $M_0$ guesses an assignment one bit per step, writes it on
a tape, and checks every clause after sorting the literal occurrences by
variable index; it runs in time $T_0=L(\log L)^{O(1)}$, and some choice
string is accepted exactly when the formula is satisfiable.

For~(2), let $M_1$ flip a coin $b$. If $b=0$ it runs $M_0$ on $T_0$ further
coins and postselects and accepts iff $M_0$ accepts. If $b=1$ it flips
$T_0+2$ further coins and postselects and rejects iff all of them are $0$.
Then $p_{\mathrm{post}}\ge2^{-(T_0+3)}>0$. On an unsatisfiable formula
$p_{\mathrm{acc}}=0$. On a satisfiable one
$p_{\mathrm{acc}}\ge2^{-(T_0+1)}$ and $p_{\mathrm{post}}-p_{\mathrm{acc}}=
2^{-(T_0+3)}$, so the conditional acceptance probability is at least
$\frac45$. So 3SAT is in $\PostBPTIME[O(T_0)]$.

For~(3), a 3SAT formula of size $s$ on $n$ variables is a fan-in-$2$
circuit $C_\varphi$ with $O(s)$ gates and description length $O(s\log s)$.
For postselection use coins $b$, $y\in\bits^n$ and $z\in\bits^{n+2}$,
postselect on $(b=0\wedge C_\varphi(y))\vee(b=1\wedge z=0)$ and accept on
$b=0\wedge C_\varphi(y)$; this has $O(s)$ gates, and the calculation above
gives conditional acceptance $0$ or at least $\frac45$. So 3SAT is decided
in space $O(s^{1-\varepsilon})$, with no logarithmic loss. Under~(1)
or~(2), 3SAT of size $s$ is decided in deterministic space
$O(T_0^{1-\varepsilon})=O(s^{1-\varepsilon'})$ for any
$\varepsilon'<\varepsilon$. In all three cases it is decided in deterministic time
$2^{O(s^{1-\varepsilon'})}$, since a halting space-$S$ machine with
$S\ge\log L$ has at most $2^{O(S)}$ configurations. The Weak Exponential Time
Hypothesis rules this out by~\cite{MIT26}.
\end{proof}

So exact counting, majority, postselection and quantum Merlin--Arthur time
with linear witnesses are all confined to space $T^{1-o(1)}$ under the Weak
Exponential Time Hypothesis, and Theorems~\ref{thm:main}
and~\ref{thm:quantum-main} are within a $\sqrt{\log T}$-type factor of what
such methods can give. A bound of the form
$\BPTIME[t]\subseteq\DSPACE(t^{1-\varepsilon})$ or
$\BQTIME[t]\subseteq\DSPACE(t^{1-\varepsilon})$, and in particular any
progress toward Williams's $\sqrt{t\log t}$ for bounded-error machines, must
use the bounded-error promise and the absence of postselection. Two natural
ways around this, enumerating the seeds of a generator and counting with
fresh coins on Williams's tree, meet the barriers discussed
in~\cite[Section~6]{BDone}.

\section*{Use of AI tools}

Claude (Anthropic),  was  used to check proofs, find counterexmaples to previous versions,  and to search the
literature, also for understanding the ETH proof. Google neon was used to check the writing.
The author is responsible for all claims.

\end{document}